
\input phyzzx
\PHYSREV
\hbox to 6.5truein{\hfill UPR-0559T }
\hbox to 6.5truein{\hfill March, 1993}
\title{RE-EXAMINATION OF GENERATION OF BARYON AND LEPTON NUMBER
ASYMMETRIES BY HEAVY PARTICLE DECAY}
\author{Jiang Liu and Gino Segr\` e}
\address{Department of Physics, University of Pennsylvania, Philadelphia,
         PA 19104}
$$ $$
\abstract
It is shown that
wave function renormalization can introduce an important
contribution to the generation of baryon and lepton number
asymmetries by heavy particle decay.  These terms, omitted in
previous analyses, are
of the same order
of magnitude as the standard terms.
A complete cancellation of leading terms
can result in some interesting cases.

\endpage

The three key elements for baryogenesis, namely baryon number violation,
$C$ and $CP$ violation and departure from thermal equilibrium were
clearly identified in Sakharov's historic paper\Ref\Sokharov{
     A. Sakharov, JETP Letters 5, 24 (1967).}
in 1967.  Realistic calculations of baryogenesis only became  possible
however in the late 1970's, after the introduction of
grand unified theories (GUTS), which provided a clear field
theoretical model in which baryon number violation occurs\rlap.
\Ref\KT{For a review of early papers on baryogenesis see for example
     E. W. Kolb and M. S. Turner, Ann. Rev. Nucl. Part. Sci. 34, 1 (1984).}

The early calculations followed a standard pattern, colloquially referred to
as the ``drift and decay mechanism''.  Pre-existing asymmetries were
presumed to be erased before the breaking of the GUT symmetry.  A particle
$S$, usually a colored Higgs boson, has a long enough lifetime so that it
is out of
thermal equilibrium when it finally decays.  Since the decaying
particle has at least two decay modes with different baryon number and
its couplings violate $CP$, the ingredients are all in place for
baryogenesis.

It was realized almost immediately that one needed to go beyond the tree
approximation in calculating decay amplitudes: otherwise $CPT$
invariance leads to a zero baryon asymmetry.  Therefore the
standard calculation involves an interference between eg. a tree
level diagram for $S$ decaying into fermions $S\to f_1f_2$ and a one
or more loop diagrams for the same process.

Many refinements and elaborations have taken place in the
past fifteen years.  Departures from the so called
``drift and decay mechanism'' have been numerous:
the most influential one has resulted from the observation by
Kuzmin, Rubakov and Shaposhnikov
that non-trivial vacuum gauge configurations can lead to a significant
baryon number violation at low temperature
\Ref\KRS{
       V. A. Kuzmin, V. A. Rubakov and M. E. Shaposhnikov, Phys. Lett.
       155B, 36 (1985).} ($\sim 100\ GeV$).
In this note, we will have nothing to say directly about low
temperature baryogenesis\rlap.\Ref\CKN{
For a recent review on the progress in electroweak baryogenesis see:
A. G. Cohen, D. B. Kaplan and A. E. Nelson,
       UCSD-PTH-9302, BUHEP-93-4, (to be published).}
   Our comments are most applicable
to the earlier calculations and variations thereof.

It was realized recently\Ref\JL{J. Liu,
      UPR-0558T, (1993).} that
wave function renormalization of a heavy unstable particle can
introduce important effects for $CP$-violating asymmetries.
Baryon number asymmetry is one such particularly interesting example.
We have realized that, whereas vertex corrections to the $S\to f_1f_2$ decay
were treated consistently, external line insertions
associated with wave function renormalizations  were not.  Since in
general these are of the same order of magnitude, the calculations change
substantially.  In one particular example, we will in fact show that
the vertex and external line insertions cancel: since, as we said earlier
baryon asymmetry is zero at tree level, these corrections are the
leading contributions to our process.

To be specific, consider a $B$- and $CP$-violating interaction
(the standard $SU(5)$ GUT model has two additional interactions
of similar form which we have omitted for simplicity)
$${\cal{L}_I}=G_{\xi}\bar u_{R,\alpha}e^c_{R}S_{\xi,\alpha}
             +F_{\xi}\bar u^c_{R,\alpha}d_{R,\beta}S_{\xi,\gamma}
             \epsilon^{\alpha\beta\gamma}+h.c.,\eqno(1)$$
where $S_{\xi,\alpha}$ is a heavy scalar belonging to the $5$ representation
of $SU(5)$. $u,  d$ and $e^c$ are the
charged fermions of the first
generation, $\alpha,\beta$ and $\gamma$ are the color indices,
and $\xi=1,2,...$ labels different species of $S$.  Complex
couplings $G_{\xi}$ and $F_{\xi}$ are the sources of $CP$ violation.
For simplicity we neglect fermion mixings.

Evidently, baryon number asymmetries generated from $S_{\xi}$ decays
are determined by the partial rate difference
$$\eqalign{\Delta_{S_{\xi}}&=\Gamma(S_{\xi}\to e\bar u^c)-
                            \Gamma(\bar S_{\xi}\to \bar eu^c),\cr
 &                          =\Gamma(\bar S_{\xi}\to d\bar u^c)
                           -\Gamma(S_{\xi}\to \bar du^c),\cr}\eqno(2)$$
where $\bar S_{\xi}, \bar e$ and $\bar u^c$ are
the $CP$ conjugates of $S_{\xi}$, $e$ and $u^c$ respectively,
and the last step of Eq. (2) follows from $CPT$.
Unless necessary, henceforth we will not display the color indices
explicitly.

It follows from Eq. (1)
that $S_{\xi}$ has only two decay modes with final states
$ e\bar u^c$ and $\bar d u^c$.
Adjoining the one-shell $t$-channel
final-state scattering $\bar du^c\to e\bar u^c$ to
$S_{\xi}\to \bar du^c$ (Fig. 1b) corresponds to a calculation of
an absorptive part of a vertex correction\Ref\LW{For a recent discussion
of final-state interactions see: L. Wolfenstein, Phys. Rev. D43, 1 (1991).}
 (Fig. 1c).  The
interference of the vertex correction (Fig. 1c) with the tree-level
amplitude (Fig. 1a) yields the standard result
$$\Delta_{S_{\xi}}(vertex)={M_{\xi}\over 32\pi^2}\sum_{\xi'}
  Im(G_{\xi}^*G_{\xi'}F_{\xi}F^*_{\xi'})\Bigl[
  1-{M_{\xi'}^2\over M_{\xi}^2}\ln\Bigl(1+{M_{\xi}^2\over M_{\xi'}^2}\Bigr)
  \Bigr],\eqno(3)$$
where $M_{\xi}$ is the mass of $S_{\xi}$.  All fermions are
massless at the scale of $M_{\xi}$.

In addition to the $t$-channel scattering, the two final states
are also related by an $s$-channel interaction.  Adjoining the
on-shell $s$-channel amplitude
$\bar du^c\to e\bar u^c$ to $S_{\xi}\to \bar du^c$ corresponds
to a calculation of an absorptive part of a wave function renormalization
correction (Fig. 1d).  If the scalars are not degenerate, the
calculation is very simple with the result from the interference of
Figs. (1a) and (1d) given by
$$\Delta_{S_{\xi}}(wave)=-{M_{\xi}\over 32\pi^2}\sum_{\xi'}
    Im(G_{\xi}^*G_{\xi'}F^*_{\xi'}F_{\xi}) {M_{\xi}^2\over
    M_{\xi}^2-M_{\xi'}^2}.\eqno(4)$$
In obtaining this result we have assumed for simplicity that
$(M_{\xi}-M_{\xi'})^2\gg (\Gamma_{\xi}-\Gamma_{\xi'})^2$,
where $\Gamma_{\xi}$ is the width of $S_{\xi}$.  This
contribution, which is of the same order as $\Delta_{S_{\xi}}(vertex)$,
has been missed by early calculations.

The significance of $\Delta_{S_{\xi}}(wave)$ may be illustrated
by its limiting values.  When $M_{\xi'}\gg M_{\xi}$ we
have from Eqs. (3) and (4)
$$\Delta_{S_{\xi}}(wave)=2\Delta_{S_{\xi}}(vertex).\eqno(5)$$
Thus, neglecting $\Delta_{S_{\xi}}(wave)$ under-estimates
$\Delta_{S_{\xi}}(total)$ by a factor of $3$ in this limit.
In the opposite limit, i.e., $M_{\xi'}\ll M_{\xi}$, Eqs. (3) and (4)
lead to
$$\Delta_{S_{\xi}}(wave)=-\Delta_{S_{\xi}}(vertex),\eqno(6)$$
and the total result even cancels to leading order in
$(M^2_{\xi'}/M^2_{\xi})$.

The relative size
and sign  of $\Delta_{S_{\xi}}(vertex)$
and $\Delta_{S_{\xi}}(wave)$  in the limit $M_{\xi'}\gg M_{\xi}$ can be
understood easily by a Fierz transformation.
In general, their final-state scattering amplitude
in  Figs. (1c) and (1d) is given by
$$\eqalign{{\cal{A}}(\bar du^c\to e\bar u^c)=-iG_{\xi'}F^*_{\xi'}
\epsilon^{\alpha\beta\gamma}&\Bigl[
{[\bar u_e(p_1)Lv_{\gamma}(k_1)][\bar v_{\beta}(k_2)Lu_{\alpha}(p_2)]\over
  (p_1-k_1)^2-M_{\xi'}^2}\cr
&+{[\bar u_e(p_1)Lu_{\alpha}(p_2)]
   [\bar v_{\beta}(k_2)Lv_{\gamma}(k_1)]\over
    (p_1+p_2)^2-M_{\xi'}^2}\Bigr],\cr}\eqno(7)$$
where the $u's$ and $v's$ are the standard Dirac spinors.
The first term arises from the $t$-channel (Fig. 1c) and
the second is due to the $s$-channel (Fig. 1d). Averaging
over the incident momenta in the center of mass frame
yields for the $J=0$ partial wave ($J$ is the total angular momentum)
$$\int^1_{-1}d\cos\theta\sum_{spin}{\cal{A}}=
C^{\alpha}[\bar u_e(p_1)Lu_{\alpha}(p_2)]
\Bigl[\Bigl[1-{M_{\xi'}^2\over s}\ln\Bigl(1+{s\over M_{\xi'}^2}\Bigr)\Bigr]
-{s\over s-M_{\xi'}^2}\Bigr],\eqno(8)$$
where $s=(k_1+k_2)^2=M_{\xi}^2$ and
$C^{\alpha}$ is an overall factor determined by the
couplings $G_{\xi'}$ and $F_{\xi'}$ and the normalization constants of
the spinors.  The ratio between the first and second terms
is precisely $\Delta_{S_{\xi}}(vertex)/\Delta_{S_{\xi}}(wave)$.

Including fermion mixings the couplings $G_{\xi}$ and $F_{\xi}$ become
matrices in flavor space.  Hence,
generally speaking,  $\Delta_{S_{\xi}}(vertex)$
and $\Delta_{S_{\xi}}(wave)$
are not simply related as in
Eqs. (5) and (6):
for vertex corrections one will have a trace
over a family matrix to the fourth power while
the wave function correction will have a product
of two traces of the square of family matrices.
Still, the order-of-magnitudes
of $\Delta_{S_{\xi}}(vertex)$ and $\Delta_{S_{\xi}}(wave)$ are the same.

The model of baryogenesis we have considered requires more than
one Higgs color triplet $(S_{\xi}\ne S_{\xi'})$ and does
not have natural flavor conservation, i.e., both
$S_{\xi}$ and $S_{\xi'}$  couple to the two scalar fermion currents.
The additional diagrams we have been discussing
will not make any contribution without these features.

On the other hand the vertex diagrams also give vanishing contributions
to baryogenesis in lowest order for the minimal model;
the first non-vanishing contribution arises
          from a three loop diagram\rlap.
          \Ref\NW{
          D. V. Nanopoulos and S. Weinberg, Phys. D20, 2484 (1979);
          S. Barr, and G. Segr\`e and
          H. A. Weldon,  Phys. Rev. D20, 2494 (1979);
          A. Yildiz and P. Cox, Phys. Rev. D21, 306 (1980);\splitout
          T. Yanagida and M. Yoshimara, Nucl. Phys. B168, 534 (1980);
          J. Ellis, M. K. Gaillard and D. V. Nanopoulos, Phys. Lett.
          80B, 360 (1979), {\it ibid} 82B, 464(E) (1979).  }
We have displayed
here the simplest $SU(5)$ like model which contributes to
baryogenesis in lowest order.

It is interesting to notice that Fig. (1d) is a
one-particle-reducible (OPR) diagram.
Even though
one can introduce a renormalization scheme in which
the renormalized self-energy matrix $\Sigma^{(R)}_{\xi\xi'}(p)$
vanishes  on-shell, i.e.,
$\Sigma^{(R)}_{\xi\xi'}(p)\vert_{p^2=M_{\xi}^2}=
\Sigma^{(R)}_{\xi\xi'}(p)\vert_{p^2=M_{\xi'}^2}=0$,
that $\Delta_{S_{\xi}}(wave)\ne 0$ is
because the kinetic energy part of the renormalized
lagrangian will not
have the standard normalization, and
the renormalized
field does not conjugate to its hermitian conjugate ( Ref. \JL\ ),
due to
the non-hermiticity of the renormalized effective lagrangian.

It seems that the situation of most interest (Ref. \KRS\ ) is that in which
the heavy particle masses  are nearly degenerate, i.e.,
$(M_{\xi}-M_{\xi'})-i(\Gamma_{\xi}-\Gamma_{\xi'})/2\to 0$.
In that case Eq. (4) is invalid.  If $CP$ violation still  can
be treated perturbatively, $\Delta_{S_{\xi}}(wave)$
can be obtained by studying the renormalization effect on
unstable particle propagator
$\langle 0\vert TS_{\xi}(x)S^{\dag}_{\xi'}(y)\vert 0\rangle$.
An analogous example with $\xi=\xi'=1$ for
the $CP$-violating partial rate difference  of the decay
$t\to bW^+, bH^+$
is discussed in Ref. \JL.\  Methods
useful for degenerate unstable particles with large
$CP$-violating interactions are still unfortunately unavailable.

Wave function renormalization also plays an important role in leptogenesis.
Consider the generation of lepton number asymmetries
by heavy Majorana neutrino decay\rlap.\REFS\YY{
         For an early discussion of baryogenesis by heavy neutrino
         decay see e.g. T. Yanagida and M. Yoshimara, Phys. Rev. D23,
         2048 (1981);  R. Barbieri, D. V. Nanopoulos and A. Masiero,
         Phys. Lett. B98, 191 (1981).}\REFSCON\FY{
         The idea that baryogenesis could be due to leptongenesis
         by heavy neutrino decay, followed by $B+L$ violation at
         the electroweak scale was proposed by
         M. Fukugita, and T. Yanagida, Phys. Lett. B174, 45 (1986);
         P. Langacker, R. D. Peccei and T. Yanagida, Mod. Phys. Lett.
         A1, 541 (1986).}\REFSCON\FYL{
         More recent analyses of the ideas proposed in Ref. \FY\
         are continued in:  M. Fukugita and T. Yanagida, Phys. Rev.
         D42, 1285 (1990);
         M. A. Luty, Phys. Rev. D45, 455 (1992).}\REFSCON\MZ{
         For some related works on baryogenesis by neutrino
         decay see e.g.
         R. N. Mohapatra, and X. Zhang, Phys. Rev. D46, 5331 (1992);
         B. Campbell, S. Davidson and K. A. Olive, `` Inflation,
         Neutrino Baryogenesis ...'', Alberta-Minnesote preprint 11/92.}
         \REFSCON\Peccei{  For a recent overall review of the
         subject see:
         R. D. Peccei, UCLA preprint,  UCLA/92/TEP/33,
         to appear in Proceedings of XXVI Int. Conf. on High Energy
         Physics.}\refsend
We will assume the neutrions get their masses by the usual
``seesaw'' mechanism
\Ref\ss{M. Gell-Mann, P. Ramond, and R. Slansky,
   in {\it Supergravity,}
   ed. P. Van Nieuwenhuizen and D. Freedman (North-Holland, Amsterdam,
   1979);  T. Yanagida, in {\it Proceedings of the Workshop on
   Unified Theories and Baryon Number in the Universe,}
   ed. A. Sawada and A. Sugamoto (KEK Report No. 79-18, Tsukuba, 1979);
   \splitout
   R. Mohapatra and G. Senjanovic, Phys. Rev. Lett. 44, 912 (1980).}
so that we have very heavy Majoranan neutrinos $N_a$ coupled to,
on the $\sim 100\ GeV$ scale, effectively massless neutrinos $\nu_b$.
The coupling between $N_a$ and $\nu_b$ is of the form
$${\cal{L}}_I=\bar N_a(V_{ab}R+V_{ba}^*L)\nu_b\phi+h.c.,\eqno(9)$$
where $\phi$ is a neutral scalar meson.  The indices $b(a)$ runs from
$1(n+1)$ to $n(2n)$,
where $n$ is the number of neutrino families, generally taken
to be three.  The
Majorana form of the mass matrix imposes the condition
$V_{ab}=V_{ba}$.
Although, strictly speaking,   Majorana neutrinos do not carry a lepton number,
$CP$ violation
can nevertheless introduce a partial rate difference
$$\Delta_{ab}=\Gamma(N_a\to\phi\nu_{R,b})-\Gamma(N_a\to
             \phi\nu_{L,b}).\eqno(10)$$
A lepton number
asymmetry can therefore be generated if we assign a lepton number
$L=\pm 1$ to the left- and right-handed light neutrinos
(the latter are of course usually referred to as anti-neutrinos
).

The interference of the tree-level amplitude (Fig. 2a) and
the vertex correction (Fig. 2b) yields
$$\eqalign{\Delta_{ab}(vertex)={1\over 64\pi^2}
\sum_{c,d}&\Bigl[
m_cIm(V_{ab}V_{cb}^*V_{dc}^*V_{da})\Bigl[1-\Bigl(1+{m_c^2\over m_a^2}\Bigr)
\ln\Bigl(1+{m_a^2\over m_c^2}\Bigr)\Bigr]\cr
&+m_aIm(V_{ab}V^*_{cb}V_{cd}V^*_{ad})\Bigl[1-{m_c^2\over m_a^2}
\ln\Bigl(1+{m_a^2\over m_c^2}\Bigr)\Bigr]\Bigr],\cr}\eqno(11)$$
where $m_a$ and $m_c$ are respectively the masses of $N_a$
and $N_c$, and
for simplicity we have neglected
scalar masses. The light neutrino masses are also neglected
since they are, for all practical purposes, massless.
The first term in Eq. (11) corresponds to
an internal mass insertion.  The second, which has
not been included in early studies, is due to an external
neutrino mass insertion.
The final-state interaction in
this vertex correction goes through a $t$-channel with
a  total angular momentum $J=1/2$.

Once again wave function renormalization gives a contribution (Fig. 2c)
of the same order as $\Delta_{ab}(vertex)$.
For non-degenerate heavy neutrinos we find
$$\Delta_{ab}(wave)={1\over 128\pi^2}
\sum_{c,d}{m_a^2\over m_a^2-m_c^2}
\Bigl[m_c Im(V_{ab}V_{cb}^*V_{dc}^*V_{da})+
      m_a Im(V_{ab}V^*_{cb}V_{cd}V_{ad}^*)\Bigr],\eqno(12)$$
where particle widths have been neglected for simplicity.  Here
the final-state interaction goes through an $s$-channel with
$J=1/2$.  Asymptotically, $\Delta_{ab}(vertex)$
and $\Delta_{ab}(wave)$ have the following simple relations
$$\eqalign{&\Delta_{ab}(wave)=\Delta_{ab}(vertex),\ \ \  {m_c\over m_a}\gg 1,
\cr
&\Delta_{ab}(wave)={1\over 2}\Delta_{ab}(vertex),\ \ \  {m_c\over m_a}\ll
1.\cr}
\eqno(13)$$
Thus, neglecting $\Delta_{ab}(wave)$ will under-estimate
the lepton number asymmetry by a factor of $2 (3/2)$ in the limit
$m_c/m_a\gg 1 (m_c/m_a\ll 1)$.

In conclusion, we have re-examined the generation of
baryon and lepton number asymmetries by heavy particle decays.
We have shown that an important piece of contribution due
to wave function renormalization has been missed by early
investigations.  This missing piece is generally  of the same order of
magnitude as the other terms calculated before, and
can result, in some interesting cases, in a complete cancellation
of the leading terms.

Current thinking often emphasizes the erasure of previously
existing baryon and lepton asymmetries at temperatures of
$\sim 100 GeV$, which tends to de-emphasize the importance
of high temperature phenomena.  We do wish to remind the reader,
nevertheless, that sphaleron processes (Ref. \CKN\ ) conserve the anomaly
$B-L$ so that earlier asymmetries may still be important\rlap.\Ref\HT{
        For a discussion of the resulting baryon number see:
        J. Harvey and M. S. Turner, Phys. Rev. D42, 3344 (1990);
        A. Nelson and S. Barr, Phys. Lett. B246, 141 (1990), and
        S. Yu, Khlebnika and M. E. Shaposhnikov, Nucl. Phys.
        B308, 835 (1988).}

\endpage
\ack
We wish to thank P. Langacker for interesting discussions.
This work was supported in part by an $SSC$ fellowship award
(J.L.) from Texas National Research Laboratory Commission,  and
by the US Department of Energy (G.S.) under the contract
DE-AC02-76-ERO-3071.
\endpage
\centerline{FIGURE CAPTIONS}

Fig. 1.  Feynman diagrams for the generation of
baryon number asymmetries from the decay of
a colored heavy scalar $S_{\xi,\alpha}$ in an $SU(5)$ model.

Fig. 2.  Feynman diagrams for the generation of
lepton number asymmetries from the decay of a
heavy Majorana neutrino $N_a$.
\endpage
\refout
\end